\documentclass[conference]{IEEEtran}
\IEEEoverridecommandlockouts

\usepackage{cite}
\usepackage{amsmath,amssymb,amsfonts}
\usepackage{graphicx}
\usepackage{textcomp}
\usepackage{xcolor,colortbl}
\usepackage{booktabs}
\usepackage{url}
\usepackage{hyperref}

\def\BibTeX{{\rm B\kern-.05em{\sc i\kern-.025em b}\kern-.08em
    T\kern-.1667em\lower.7ex\hbox{E}\kern-.125emX}}

\hypersetup{hidelinks}
\newif\ifanonymous
\anonymousfalse

\begin{document}

\title{Multiplexing Neural Audio Watermarks\\ with Adaptive Routing}

\ifanonymous
\author{\IEEEauthorblockN{Anonymous Authors}
\IEEEauthorblockA{\phantom{Anonymous Institution}\\
\phantom{Anonymous Country}\\
\phantom{anonymous@example.com}}}
\else
\author{\IEEEauthorblockN{Zheqi Yuan\IEEEauthorrefmark{1}\quad
Yucheng Huang\IEEEauthorrefmark{1}\quad
Guangzhi Sun\IEEEauthorrefmark{2}\quad
Zengrui Jin\IEEEauthorrefmark{1}\quad
Chao Zhang\IEEEauthorrefmark{1}}
\IEEEauthorblockA{\IEEEauthorrefmark{1}Tsinghua University, China\qquad
\IEEEauthorrefmark{2}University of Cambridge, United Kingdom\\
yuanzq22@mails.tsinghua.edu.cn,\quad huang-yc22@mails.tsinghua.edu.cn\\
gs534@cam.ac.uk,\quad zrjin@tsinghua.edu.cn,\quad cz277@tsinghua.edu.cn}}
\fi

\maketitle

\begin{abstract}
Audio watermarking supports speech authenticity verification. We study whether combining heterogeneous watermarks can retain at least one detectable provenance signal when their failure modes differ. This Any-survival objective concerns complementary evidence retention, not simultaneous survival of all constituent marks or recovery of all payloads. To our knowledge, this is the first systematic study of neural audio watermark multiplexing, with a scoped benchmark covering five released systems, parallel and sequential baselines, and 14 evaluation conditions. The benchmark shows complementary failure modes, but naive composition does not reliably turn them into system-level robustness under the Any-survival objective. We therefore formulate multiplexing as watermark allocation and study perceptual-adaptive time-frequency multiplexing (PA-TFM), a training-free routing method, and MaskNet, a learned time-domain router for separately trained systems with native detectors. We use the five-system scoped benchmark to characterize multiplexing behavior and the AudioSeal–PerTh pair as a representative heterogeneous pair for adaptive watermark routing. Compared with direct parallel composition, MaskNet improves average TPR@1\%FPR from 0.75 to 0.88 and SNR from 15.20 dB to 25.36 dB; compared with PA-TFM, it gives a clear system-level robustness gain under the Any-survival objective while retaining similar high-fidelity behavior. A SpeechTokenizer-aware case study further improves average TPR@1\%FPR to 0.91 and SpeechTokenizer robustness from 0.20 to 0.60, showing that channel-adapted constituents can enter the same routing framework.\footnote{\raggedright Our code can be found at \url{https://github.com/eer000/Multiplexing-Neural-Audio-Watermarks-with-Adaptive-Routing}.}
\end{abstract}


\section{Introduction}
Audio watermarking embeds an imperceptible signal into speech to support provenance tracking, ownership claims, or synthetic-speech verification. Neural systems are commonly evaluated as single watermarks under signal and codec transformations \cite{AudioSeal2024,WavMark2023,Singh2024SilentCipher,Liu2024TimbreWatermarking}. 
This single-watermark view has two limitations. First, distribution chains may involve multiple model providers, platforms, and downstream applications requiring independently verifiable provenance evidence in the same audio. Second, no existing watermark is uniformly robust: systems that survive one class of transformations may fail under another. If heterogeneous watermarks have complementary failure modes, a multiplexed waveform could retain verifiable evidence even when one constituent is weakened or removed. The multi-stakeholder setting motivates studying shared waveforms, but our evaluated question is narrower: can routing independently developed watermarks improve the probability that at least one native detector verifies the waveform after transformation, while controlling embedding distortion? We evaluate multiplexing under an Any-survival objective: a waveform is verifiable if at least one native constituent detector accepts it under a shared low-false-positive budget. This matches provenance-retention scenarios where independent evidence can be redundant, but is weaker than requiring all payloads or all constituent marks to survive simultaneously.

This setting is distinct from making a single watermark stronger. When several neural watermarks are combined, added watermarks may interfere, sequential embedding may overwrite earlier structures, and extra energy may degrade perceptual quality. The robustness benefit is therefore not trivial: it depends on whether composition improves system-level Any-survival. A system-level gain alone does not establish retention of every stakeholder's evidence. We call this neural audio watermark multiplexing: combining independently detectable neural watermarks in one waveform while controlling distortion and retaining robustness across diverse transformations.

In this paper, we conduct, to our knowledge, the first systematic investigation of neural audio watermark multiplexing and introduce a scoped benchmark covering 5 released systems, their direct parallel and sequential combinations, and 14 evaluation conditions covering no attack and signal, codec, neural reconstruction, reverberation, and targeted transformations. The benchmark shows complementary failure modes, but naive combination strategies, such as direct addition and sequential embedding, do not reliably convert this complementarity into stable system-level Any-survival robustness. Therefore, we propose perceptual-adaptive time-frequency multiplexing (PA-TFM), a training-free method that allocates watermarks using perceptual time-frequency masks. We further propose MaskNet, a learned time-domain router that predicts carrier-dependent masks for frozen constituent watermarks.

On the representative AudioSeal–PerTh pair, MaskNet improves average detection and fidelity over direct stacking. To address the shared SpeechTokenizer failure mode, we adapt AudioSeal through a differentiable proxy; routing the adapted constituent further improves average and tokenizer-specific robustness.
Our contributions are summarized as follows.
\begin{itemize}
  \item We formulate neural audio watermark multiplexing as allocation among independently trained systems with native detectors, and benchmark 5 released systems and direct compositions under 14 evaluation conditions.
  \item We identify complementary failures not reliably exploited by naive composition, and propose training-free perceptual time-frequency routing (PA-TFM) and learned carrier-dependent time-domain routing (MaskNet).
  \item We demonstrate improved robustness–fidelity trade-offs on AudioSeal–PerTh and show that the same routing framework can incorporate a tokenizer-adapted constituent to mitigate a shared reconstruction failure.
\end{itemize}

\section{Related Work}
Traditional audio watermarking embeds structured signals through spread spectrum or quantization index modulation \cite{Boney1996,Chen2001QIM}, whereas neural watermarking learns embedding and detection jointly. Representative post-hoc systems include WavMark, AudioSeal, PerTh, MaskMark, IDEAW, SilentCipher, and Timbre Watermarking \cite{WavMark2023,AudioSeal2024,Perth2023,OReilly2024MaskMark,Li2024IDEAW,Singh2024SilentCipher,Liu2024TimbreWatermarking}. They differ in architecture, perceptual constraints, and detector interface, ranging from presence scores to synchronized messages or message confidences. Robustness evaluation has also become systematic: AudioMarkBench studies watermark removal and forgery under no-box, black-box, and white-box perturbations \cite{NEURIPS2024_AudioMarkBench}, RAW-Bench exposes representation-dependent gaps across audio domains and learned compression \cite{ozer25_interspeech}, and broader analyses organize signal-level, physical, neural, and adaptive threats \cite{wen2025sokrobustaudiowatermarking}. Comparative studies also distinguish proactive watermarking from passive deepfake detection \cite{wu25i_interspeech}. Recent work further shows that post-hoc marks can be shallow under realistic transformations or overwritten by another embedder \cite{o'reilly2025deep,yao2025mineoverwritingattacksneural}. Specialized neural designs further study jointly trained speech embedders, rerecording and desynchronization resilience, key-dependent recovery, cross-attention decoding, and discrete intermediate representations \cite{Pavlovic2022DNNWatermark,Liu2023DeAR,Li2024DRAWDR,Xu2025WAKE,Liu2025XAttnMark,Ji2024DiscreteWM}. These studies, however, still evaluate one embedding strategy at a time.

Another line of work embeds provenance into generation or coding pipelines. Latent, diffusion, and parameter-level watermarks bind the mark to generative representations or model weights \cite{Roman2024LatentWO,Liu2024GROOTGR,Ren2025P2MarkPP}, while TraceableSpeech, HiFi-GANw, TriniMark, and DualMark integrate watermarking into TTS, vocoder, model-instance, or data-origin tracing pipelines \cite{Zhou2024TraceableSpeechTP,Cheng2024HiFiGANwWS,Li2025TriniMarkAR,Yang2025DualMarkIM}. Related pipeline-level variants include attributable speech generation, collaborative synthesis, multimodal localization, and efficient distillation for scalable deployment \cite{Cho2022AttributableWO,Juvela2023CollaborativeWF,Zhou2024V2AMark,Cui2025EfficientSW}. Codec-aware methods such as collaborative codec augmentation and WMCodec improve robustness when the synthesis or codec pipeline is controllable \cite{Juvela2024AudioCA,Zhou2024WMCodecEN}. Learned reconstruction channels, including EnCodec, DAC, and SpeechTokenizer, are challenging because neural latent or RVQ representations can discard low-energy watermarks \cite{Defossez2022HighFN,NEURIPS2023_DAC,2024speechtokenizer}. Our setting is different: frozen, independently deployed watermarking systems must share one waveform while retaining native detectors, without a universal decoder or joint retraining. This creates interference, overwriting, and distortion trade-offs that are absent when strengthening a single watermark.

\section{Methodology}
\subsection{Experimental Framework and Metric Definitions}
Let $x\in\mathbb{R}^{L}$ be a clean waveform and let $\mathcal{W}_i(x)=x+\delta_i$ denote the output of a frozen watermark embedder. We call $\delta_i=\mathcal{W}_i(x)-x$ the watermark added by that embedder; this is an input-output definition and does not assume that the released system explicitly exposes this additive component. We use A, P, S, T, and W for the five single systems. The symbol ``+'' is reserved for parallel watermark addition, while ``-'' denotes sequential application from left to right. Thus, A+P is parallel composition and A-P applies A before P. Parallel multiplexing superimposes watermarks, whereas sequential multiplexing applies complete embedders in order:
\begin{equation}
 \tilde{x}_{\rm par}=x+\sum_i\alpha_i\delta_i,\qquad
 \tilde{x}_{A-P}=\mathcal{W}_{P}(\mathcal{W}_{A}(x)).
 \label{eq:naive}
\end{equation}
Direct parallel composition uses each system's native watermark scale, \textit{i.e.} $\alpha_i=1$, rather than equal-SNR rescaling; comparability is maintained within each multiplexing comparison by using the same A/P watermarks, utterances, transformations, and frozen detectors across parallel composition, PA-TFM, and MaskNet.

Detectors provide continuous presence scores. A multiplexed sample is positive under the system-level \emph{Any} rule when at least one constituent accepts it. Single-watermark detection is the one-constituent case of the system definition below.
The principal evaluation uses 2620 paired transformed watermarked positives and clean negatives per method--condition cell. Any-survival measures evidence retention, not simultaneous payload recovery; payload-level coexistence remains a stricter objective for future work.

For method $m$ and transformation $c$, let $x_n$ be clean source $n$ and $\tilde{x}_{m,n}$ its watermarked counterpart. Every native detector receives paired inputs $z^-_{c,n}=\mathcal{T}_c(x_n)$ and $z^+_{m,c,n}=\mathcal{T}_c(\tilde{x}_{m,n})$, with the same transformation condition applied to both classes.
Let $s_i(z)$ be the continuous score of detector $i$, oriented so larger values indicate presence. For constituent set $\mathcal{I}_m$, the Any decision is
\begin{equation}
 a_m(z;\boldsymbol{\tau})=
 \mathbb{1}\!\left[\bigvee_{i\in\mathcal{I}_m}s_i(z)\geq\tau_i\right].
 \label{eq:any_explicit}
\end{equation}
The system rates average that decision over the two pools:
\begin{equation}
 \begin{aligned}
 \mathrm{TPR}_{m,c}(\boldsymbol{\tau})
 &=\frac{1}{N_+}\sum_{n=1}^{N_+}a_m(z^+_{m,c,n};\boldsymbol{\tau}),\\
 \mathrm{FPR}_{m,c}(\boldsymbol{\tau})
 &=\frac{1}{N_-}\sum_{n=1}^{N_-}a_m(z^-_{c,n};\boldsymbol{\tau}).
 \end{aligned}
\end{equation}
We report TPR@1\%FPR by jointly sweeping constituent thresholds to maximize empirical system TPR subject to system FPR $\leq0.01$. Detector $i$ uses the same $\tau_i$ for both classes; thresholds may differ across detectors. The OR false-positive rate is evaluated on paired score vectors, not separate 1\% budgets per detector. 

For constituent analysis, each detector's TPR@1\%FPR is computed separately on the same method--condition waveforms under its own false-positive constraint. These diagnostic thresholds differ from the joint Any thresholds. Neither the sum nor the maximum of these diagnostic TPRs reconstructs the system Any result. Presence detection does not measure correct payload recovery. 

\subsection{Multiplexing Methods}
\subsubsection{PA-TFM}
PA-TFM uses a hard-coded routing heuristic motivated by auditory masking, selected by parameter search rather than gradient-based training of a routing network. Let $X(t,f)=\mathcal{F}\{x\}$ and $D_i(t,f)=\mathcal{F}\{\delta_i\}$ denote the carrier and watermark spectra. Instead of assigning global weights, complementary masks $M_i(t,f)$ are derived from frame-level local signal-to-noise ratio (SNR) and spectral flatness as heuristic indicators of auditory masking capacity. The fused spectrum is computed as:
\begin{equation}
  \tilde{X}(t,f)=X(t,f)+\sum\nolimits_i M_i(t,f)D_i(t,f).
  \label{eq:patfm}
\end{equation}
The masks are smoothed before inverse STFT, making PA-TFM a training-free perceptual-routing baseline that reduces abrupt mask transitions but does not guarantee codec robustness, since perceptually masked watermarks may still be removed by lossy coding.

Here $X,D_i\in\mathbb{C}^{T\times F}$ are complex STFT coefficients and $M_i\in[0,1]^{T\times F}$ is real-valued. Nonnegative real masks scale constituent coefficients without changing their phases; inverse STFT uses the fused complex $\tilde{X}$. No magnitude-only reconstruction is implied.

Within predefined SNR and spectral-flatness ranges, we sweep linear mappings from A-dominant to P-dominant allocation at a fixed weight sum. Using the same training split as tokenizer adaptation, we select the mapping maximizing mean Any-rule TPR@1\%FPR over the 14 evaluation conditions, including no attack. Parameters remain fixed at test time, while frame-level weights depend on the input.

\subsubsection{MaskNet}
MaskNet replaces the fixed rule with learned time-domain routing. A lightweight 1D-CNN predicts sample-wise routing masks $m_A,m_P\in\mathbb{R}^{L}$ from the carrier waveform and constructs
\begin{equation}
  \tilde{x}=x+m_A\odot\delta_A+m_P\odot\delta_P.
  \label{eq:masknet}
\end{equation}
The A and P embedders and detectors remain frozen. Training minimizes
\begin{equation}
 \mathcal{L}=\mathcal{L}_{\rm det}+\lambda_{w}\mathcal{L}_{\rm wav}
 +\lambda_{q}\mathcal{L}_{\rm quiet}+\lambda_{r}\mathcal{L}_{\rm reg},
 \label{eq:maskloss}
\end{equation}
where detection loss is balanced against waveform distortion, quiet-region suppression, and mean mask activation, as defined below. We denote this pair-specific router as MaskNet(A,P).

For a minibatch of $B$ length-$L$ waveforms, the waveform loss is mean squared error: $\mathcal{L}_{\rm wav}=\frac{1}{BL}\sum_{b,\ell}(\tilde{x}_{b,\ell}-x_{b,\ell})^2$.
The quiet-region weight and loss are
\begin{equation}
 \begin{aligned}
 e_{b,\ell}&=\frac{1}{513}\sum_{k=-256}^{256}|x_{b,\ell+k}|,\\
 q_{b,\ell}&=\max(0,0.01-e_{b,\ell})/0.01,\\
 \mathcal{L}_{\rm quiet}
 &=\frac{1}{BL}\sum_{b,\ell}(m_{A,b,\ell}+m_{P,b,\ell})q_{b,\ell}.
 \end{aligned}
\end{equation}
The envelope uses zero padding at boundaries. The branch regularizer is
\begin{equation}
 \mathcal{L}_{\rm reg}=0.001\,\mathbb{E}_{b,\ell}[m_A]
 +0.1\,\mathbb{E}_{b,\ell}[m_P].
\end{equation}
We use $\lambda_r=1$ in Eq.~\ref{eq:maskloss}. After additive Gaussian noise, the heterogeneous detection objective is
\begin{equation}
 \mathcal{L}_{\rm det}=\lambda_\text{A}\operatorname{BCE}(p_\text{A},y_\text{A})
 +\lambda_\text{P}\bigl(1-\mathbb{E}[c_\text{P}]\bigr),
 \label{eq:detloss}
\end{equation}
where $p_\text{A}$ is AudioSeal's differentiable prediction for the message and $c_\text{P}$ is PerTh's watermark confidence. This optimizes both constituents through one router without forcing incompatible detector outputs into a shared decoder.

Discrete codecs and tokenizers are excluded from the MaskNet gradient path. Results under MP3, Opus, EnCodec, DAC, and SpeechTokenizer therefore measure system-level Any-survival beyond the Gaussian-noise training channel, not codec-specific routing. Constituent preservation is assessed separately below.

A and P expose distinct training signals and complementary attack profiles. MaskNet allocates their frozen watermarks in one pass; native detectors remain unchanged.

\begin{figure}[t]
  \centering
  \begin{minipage}{.92\linewidth}
    \centering\includegraphics[width=\linewidth]{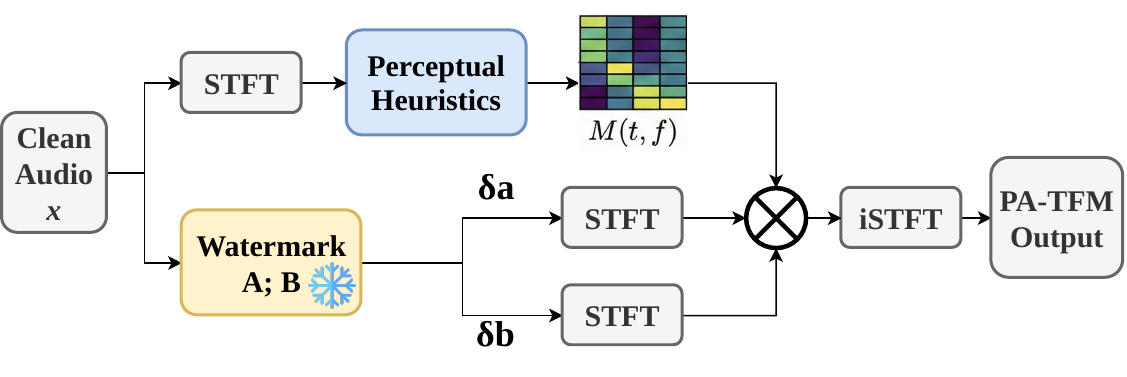}
    \centerline{(a) PA-TFM}
  \end{minipage}

  \vspace{2pt}

  \begin{minipage}{.92\linewidth}
    \centering\includegraphics[width=\linewidth]{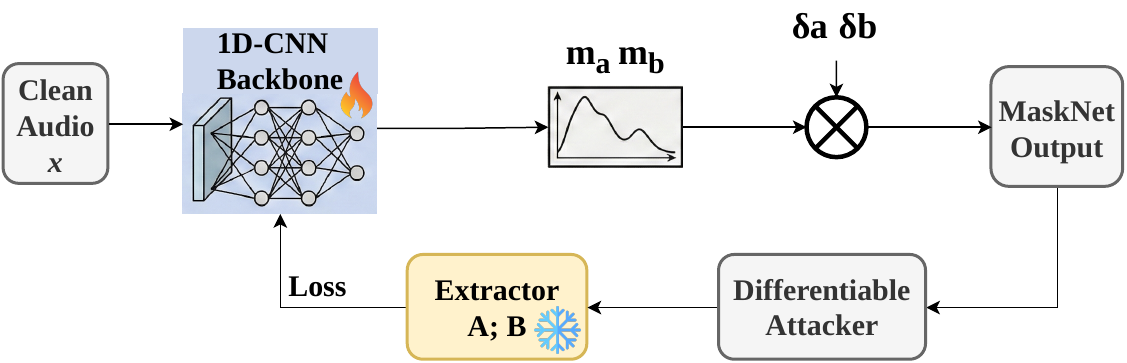}
    \centerline{(b) MaskNet}
  \end{minipage}
  \caption{Watermark-routing methods. PA-TFM generates time-frequency masks with a fixed rule; MaskNet learns time-domain masks for frozen A and P watermarks.}
  \label{fig:multiplexing}
\end{figure}

\subsection{SpeechTokenizer-Aware Case Study}
The original constituents share a weakness to SpeechTokenizer, which maps encoder features to discrete RVQ codewords and resynthesizes speech from retained tokens \cite{2024speechtokenizer}. Watermarks that do not change a selected codeword disappear, while codeword changes are non-differentiable. The evaluated routing methods still show a pronounced weakness under this channel. We therefore test whether the multiplexing framework can accept a newly adapted constituent.

We fit a 12-block gated residual ProxyNet $P_\phi$ with 96 hidden channels to the frozen SpeechTokenizer path $S(\cdot)$ using waveform, multi-resolution spectral, and detector-response objectives,
\begin{equation}
 \begin{aligned}
 \mathcal{L}_{\rm proxy}
 =&\,\lambda_1\operatorname{MAE}(P_\phi(x),S(x))\\
 &+\lambda_2\operatorname{MSE}(P_\phi(x),S(x))\\
 &+\lambda_s\mathcal{L}_{\rm spec}+\lambda_d\mathcal{L}_{\rm resp}.
 \end{aligned}
\end{equation}
After fitting on clean and watermarked speech, we freeze the proxy and AudioSeal detector, updating only the generator's decoder and message processor as defined below. The adapted constituent is $A_\text{T}$ (tokenizer-adapted, not Timbre). MaskNet($A_\text{T}$,P) is retrained with the same architecture, losses, and procedure, replacing $\delta_A$ with $\delta_{A_\text{T}}$, rather than substituted at inference. ProxyNet supports this extensibility study, not a separate contribution.

Detector-response matching encourages the proxy to reproduce watermark detector responses. During adaptation, clean-path recovery and distortion penalties protect ordinary AudioSeal performance; model selection and final evaluation use the true tokenizer, not proxy output.

Each ProxyNet block uses length-preserving, dilated kernel-7 convolutions on the 96-channel states:
\begin{equation}
 u_j=\tanh(W_{f,j}*h_j)\odot\sigma(W_{g,j}*h_j).
\end{equation}
A pointwise convolution precedes each residual update. The spectral loss compares multi-resolution log-STFT magnitudes; the response loss matches frozen AudioSeal presence and message outputs using MSE. 

For adapted watermarked audio $w$, the surrogate-gradient construction is $v=\operatorname{sg}[S(w)]+P_\phi(w)-\operatorname{sg}[P_\phi(w)]$, where $\operatorname{sg}$ stops gradients. Its forward value is $S(w)$ and its input derivative is supplied by frozen $P_\phi$. Let $p$ and $d$ be frozen detector message and presence probabilities, and $y$ the target message. The adaptation objective is
\begin{equation}
 \begin{aligned}
 \mathcal{L}_{\rm adapt}
 =&\,\ell_{\rm bit}(p(v),y)
 +\beta_p\ell_{\rm bit}(p(P_\phi(w)),y)\\
 &+\beta_c\ell_{\rm bit}(p(w),y)
 +\beta_d\operatorname{BCE}(d(v),\mathbf{1})\\
 &+\beta_w\operatorname{MSE}(w,x)
 +\beta_s[\gamma-\operatorname{SNR}(x,w)]_+.
 \end{aligned}
\end{equation}
Here $\ell_{\rm bit}$ denotes the message-recovery loss and $[\cdot]_+$ the positive part; the clean path means watermarked audio without tokenization. SNR is $10\log_{10}(\mathbb{E}[x^2]/\mathbb{E}[(w-x)^2])$ with numerical stabilization.  

\begin{table*}[!t]
\centering
\caption{Robustness under the scoped evaluation suite, reported as TPR@1\%FPR. Each system is evaluated with paired watermarked positives and transformed clean negatives. A multiplexed system is positive when any constituent detector accepts it. Overall is the unweighted mean across the 14 evaluation conditions. Higher is better.}
\label{tab:robustness}
\scriptsize
\setlength{\tabcolsep}{2.60pt}
\renewcommand{\arraystretch}{1.08}
\resizebox{0.97\textwidth}{!}{%
\begin{tabular}{lccccccccccccccc}
\toprule
Method         & NA     & GN     & UN     & ZM     & FT     & EC     & MP     & OP     & EN     & ST     & DA     & RV     & AW     & PW     & Overall \\
\midrule
AudioSeal (A)              & 1.00  & 0.78   & 1.00  & 1.00  & 0.00   & 1.00  & 1.00  & 0.38   & 1.00  & 0.01   & 1.00  & 0.75   & 0.05   & 1.00  & 0.71 \\
PerTh (P)              & 1.00  & 0.81   & 1.00  & 1.00  & \textbf{0.75}   & 0.48   & 1.00  & 1.00  & 0.18   & 0.20   & 0.74   & 0.00   & 1.00  & 0.31   & 0.68 \\
SilentCipher (S)              & 1.00  & 0.25   & 0.13   & 0.85   & 0.14   & 0.11   & 0.39   & 0.35   & 0.02   & 0.02   & 0.01   & 0.26   & 1.00  & 0.79   & 0.38 \\
Timbre (T)              & 1.00  & 0.27   & 1.00  & 1.00  & 0.05   & 1.00  & 1.00  & 0.12   & 0.45   & 0.02   & 0.79   & 0.67   & \textbf{1.00}  & 1.00  & 0.67 \\
WavMark (W)              & 1.00 & 0.15 & 0.45 & 0.37 & 0.55 & 1.00 & 1.00 & \textbf{1.00} & 0.00 & 0.00 & 0.00 & \textbf{0.91} & 0.27 & 1.00 & 0.55 \\
\midrule
Parallel A+P   & 1.00 & 0.79  & 1.00  & 1.00  & 0.37 & 0.94  & 1.00   & 0.76   & 1.00  & 0.01  & 1.00  & 0.72 & 0.40 & 0.55 & 0.75 \\
Sequential A-P & 1.00 & 0.81  & 1.00  & 0.68   & 0.70 & 0.70  & 1.00   & 0.88   & 0.21   & 0.11  & 0.61   & 0.31 & 0.93 & 0.30 & 0.66 \\
Sequential P-A & 1.00 & 0.70  & 0.95   & 0.80   & 0.02 & 1.00 & 1.00   & 0.36   & 1.00  & 0.01  & 0.84   & 0.69 & 0.09 & 0.85 & 0.67 \\
PA-TFM         & 1.00 & 0.70  & 1.00  & 1.00  & 0.42 & 0.92  & 0.89    & 0.36   & 0.70   & 0.01  & 1.00  & 0.76 & 0.87 & 0.90 & 0.75 \\
MaskNet(A,P)   & 1.00 & \textbf{0.82}  & 1.00  & 1.00  & 0.71 & 1.00 & 1.00   & 0.90   & 1.00  & 0.20  & 1.00  & 0.71 & 0.96 & 1.00 & 0.88 \\
\midrule
$A_\text{T}$          & 1.00 & 0.73 & 1.00 & 0.98  & 0.00   & 1.00  & 1.00  & 0.35 & 1.00 & \textbf{0.63} & 1.00 & 0.72 & 0.00 & 1.00 & 0.74 \\
MaskNet($A_\text{T}$,P) & \textbf{1.00} & 0.78 & \textbf{1.00} & \textbf{1.00} & 0.72   & \textbf{1.00}  & \textbf{1.00}  & 0.94 & \textbf{1.00} & 0.60 & \textbf{1.00} & 0.70 & 0.98 & \textbf{1.00} & \textbf{0.91} \\
\bottomrule
\end{tabular}}
\end{table*}

\section{Experimental Setup}
\textbf{Data and systems.}
For cross-domain evaluation, we use two datasets: the \texttt{test-clean} subset of LibriSpeech \cite{Panayotov2015} (approximately 5.5 hours) and the \texttt{en-AU} subset of Common Voice Corpus 24.0 (approximately 33.5 hours). All audio files are resampled to 16 kHz. Principal method--attack cells dynamically sample 2620 watermarked positives and 2620 clean negatives from these test sets. We evaluate A, P, S, T, and W as single-watermark references. Furthermore, the core MaskNet comparison is trained on A+P using the \texttt{train-clean} subset of LibriSpeech, ensuring strictly disjoint training and test speakers.

AudioSeal, PerTh, SilentCipher, and WavMark operate at 16~kHz. Timbre uses its released 22.05~kHz checkpoint; its waveform is converted to the common 16~kHz attack interface and returned to the native rate for detection. For every source, the clean negative and corresponding watermarked positive undergo the same transformation. The transformation can change both watermarked and clean scores; the transformed clean pool is needed to measure false positives.
The continuous score is A's clip-level watermark probability, P's non-rounded decoder confidence, S's returned confidence, soft agreement between T's decoder logits and its fixed detection pattern, and W's latent decoding confidence. We retain these outputs before hard decisions, giving the five systems one confusion-matrix protocol despite their different representations.

\textbf{Transformations.}
The scoped suite contains no attack (NA); Gaussian and uniform noise (GN/UN); zero and FFT masking (ZM/FT); echo (EC); MP3 \cite{MP3ISO1993} and Opus \cite{OpusRFC} compression (MP/OP); EnCodec, SpeechTokenizer, and DAC reconstruction (EN/ST/DA); room impulse response (RV); and targeted AudioSeal and PerTh white-box transformations (AW/PW). Parameters are fixed across methods, and both transformed positive and negative score pools enter the evaluation. 

The suite is scoped rather than exhaustive. GN/UN test additive interference; ZM/FT remove local time- or frequency-domain content; EC and RV model delayed and reverberant propagation; MP/OP test conventional distribution codecs; and EN/ST/DA test learned reconstruction. AW and PW optimize against one named detector. They measure targeted fragility and the value of independent evidence, not security against an attacker jointly optimizing every constituent.

\textbf{Metrics.}
We report TPR@1\%FPR as the detection metric, using the positive and negative score sweep defined in Section III-A. It is the highest TPR at an empirical FPR no greater than 1\%, giving every method the same system-level false-positive budget. The principal detection results are point estimates from 2620 samples per class; small per-cell differences should be read with caution, and we therefore emphasize Overall trends rather than individual-cell comparisons.
Quality before attack is measured by SNR, PESQ \cite{Rix2001}, STOI \cite{Taal2011}, WER from Whisper large-v3, and an ABX listening test. The ABX test uses 50 matched clean--watermarked pairs and 10 listeners. For each pair, A and B are the two versions in randomized order, X randomly repeats one of them, and the listener identifies whether X matches A or B. We report the fraction of correct decisions over all 500 trials; 50\% is chance performance. Attack-induced quality is kept separate from watermark embedding fidelity.

\begin{figure}[t]
  \centering
  \includegraphics[width=\columnwidth]{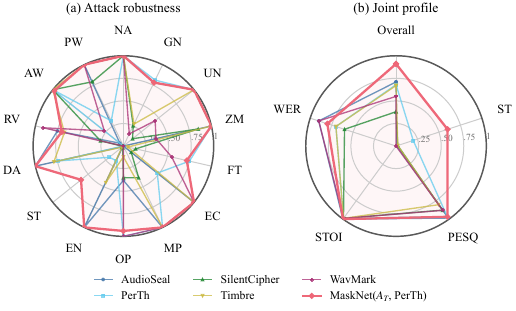}
  \caption{Radar summaries for the single-watermark baselines and MaskNet($A_\text{T}$,P). Abbreviations: NA no attack; GN/UN noise; ZM/FT masking; EC echo; MP/OP codecs; EN/ST/DA neural reconstruction; RV reverberation; AW/PW targeted white-box transformations.}
  \label{fig:radar}
\end{figure}

SNR measures perturbation energy, not perceptual quality alone; PESQ, STOI, WER, and ABX assess quality, intelligibility, recognition, and discriminability. Figure~\ref{fig:radar}(a) uses attack-wise TPRs from Table~\ref{tab:robustness}. Panel (b) retains Overall, ST, and STOI, maps PESQ as $(\mathrm{PESQ}-1)/3.5$ and WER as $1-\mathrm{WER}/0.10$, and clips values to $[0,1]$. These fixed mappings are identical across methods.

\begin{table}[!b]
\centering
\caption{Mean per-detector TPR@1\%FPR over 14 evaluation conditions. Bold: column best.}
\label{tab:constituents_cr}
\small
\begin{tabular}{lcc}
\toprule
Waveform method & A detector & P detector\\
\midrule
Parallel A+P & 0.71 & 0.67\\
Sequential A-P & 0.63 & 0.66\\
Sequential P-A & 0.67 & 0.61\\
PA-TFM & 0.70 & 0.67\\
MaskNet(A,P) & \textbf{0.76} & \textbf{0.70}\\
\bottomrule
\end{tabular}
\end{table}

\section{Results}
\subsection{Multiplexing Robustness}
Table~\ref{tab:robustness} shows complementary failure modes across the 14 evaluation conditions: A is strong under learned reconstruction but weak to frequency masking and AW, whereas P covers some of these cases but fails under reverberation and EnCodec. However, direct multiplexing does not reliably exploit this complementarity, as parallel A+P and both sequential orders remain close to the stronger single systems rather than providing a stable overall lift.

Figure~\ref{fig:radar} summarizes attack-wise coverage and the joint robustness--fidelity profile. MaskNet($A_\text{T}$,P) leads on Overall and ST, while several single systems remain competitive on individual fidelity axes.

Targeted and frequency-domain attacks expose different profiles for parallel and sequential composition, with order-dependent Any-rule results. These aggregate scores neither identify the surviving constituent nor establish overwriting; constituent-level measurements follow below.

PA-TFM improves targeted white-box performance over direct addition but loses ground elsewhere, including MP3: perceptual routing is a fixed control, not evidence of codec resistance. MaskNet(A,P) gives a stronger, more balanced profile across conventional-codec, neural-reconstruction, and targeted transformations, consistent with data-dependent routing.

MaskNet(A,P) performs well on conventional codecs and learned reconstruction absent from its training loop, but is not codec-invariant, particularly under ST. Coverage matches the complementary standalone profiles (EN/DA for A, MP/OP for P), without establishing retention of each constituent's standalone performance.

\begin{figure}[t]
  \centering
  \begin{minipage}[b]{0.48\linewidth}
    \centering\includegraphics[width=\linewidth]{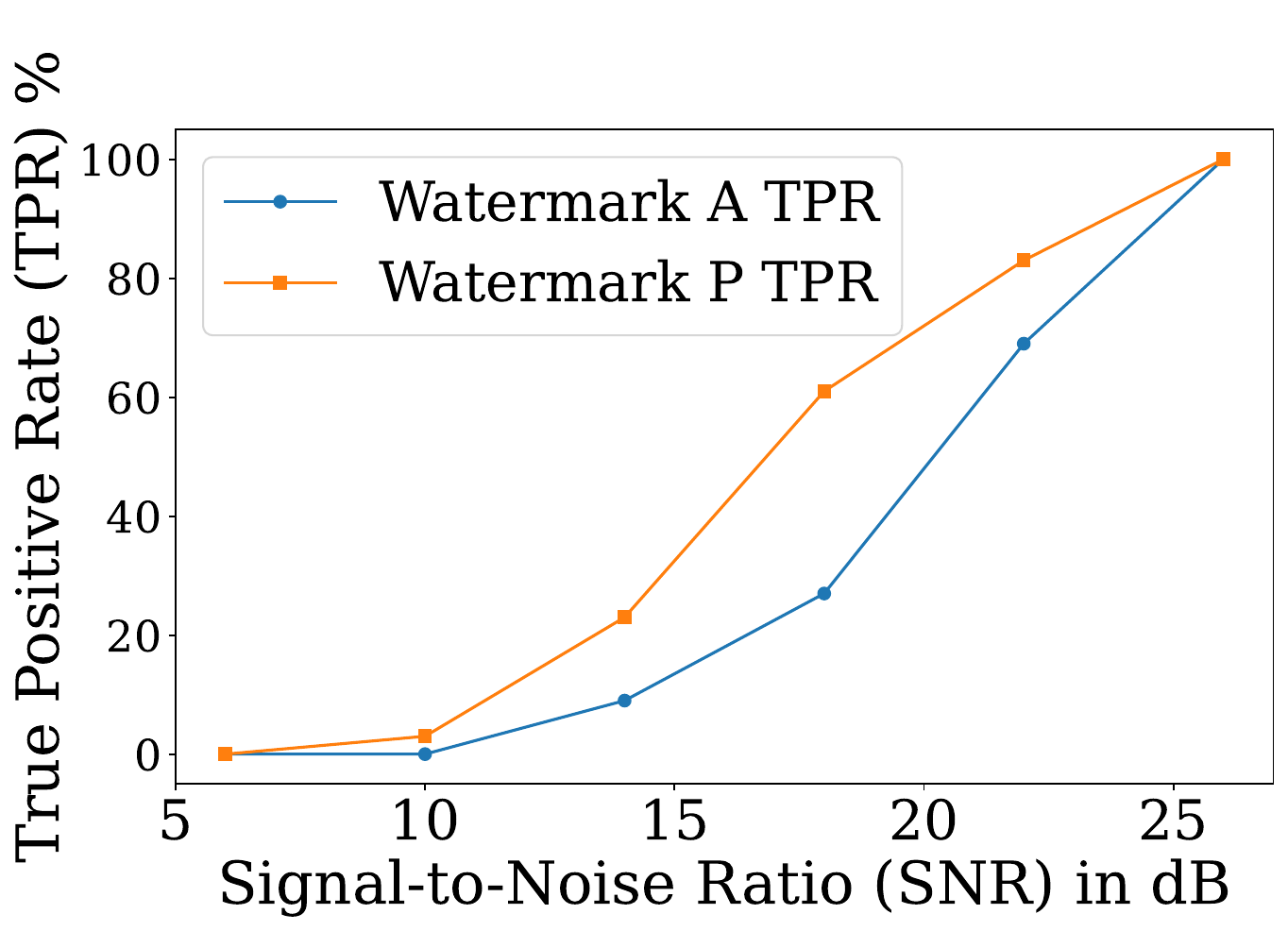}
    \centerline{(a) Gaussian noise}
  \end{minipage}\hfill
  \begin{minipage}[b]{0.48\linewidth}
    \centering\includegraphics[width=\linewidth]{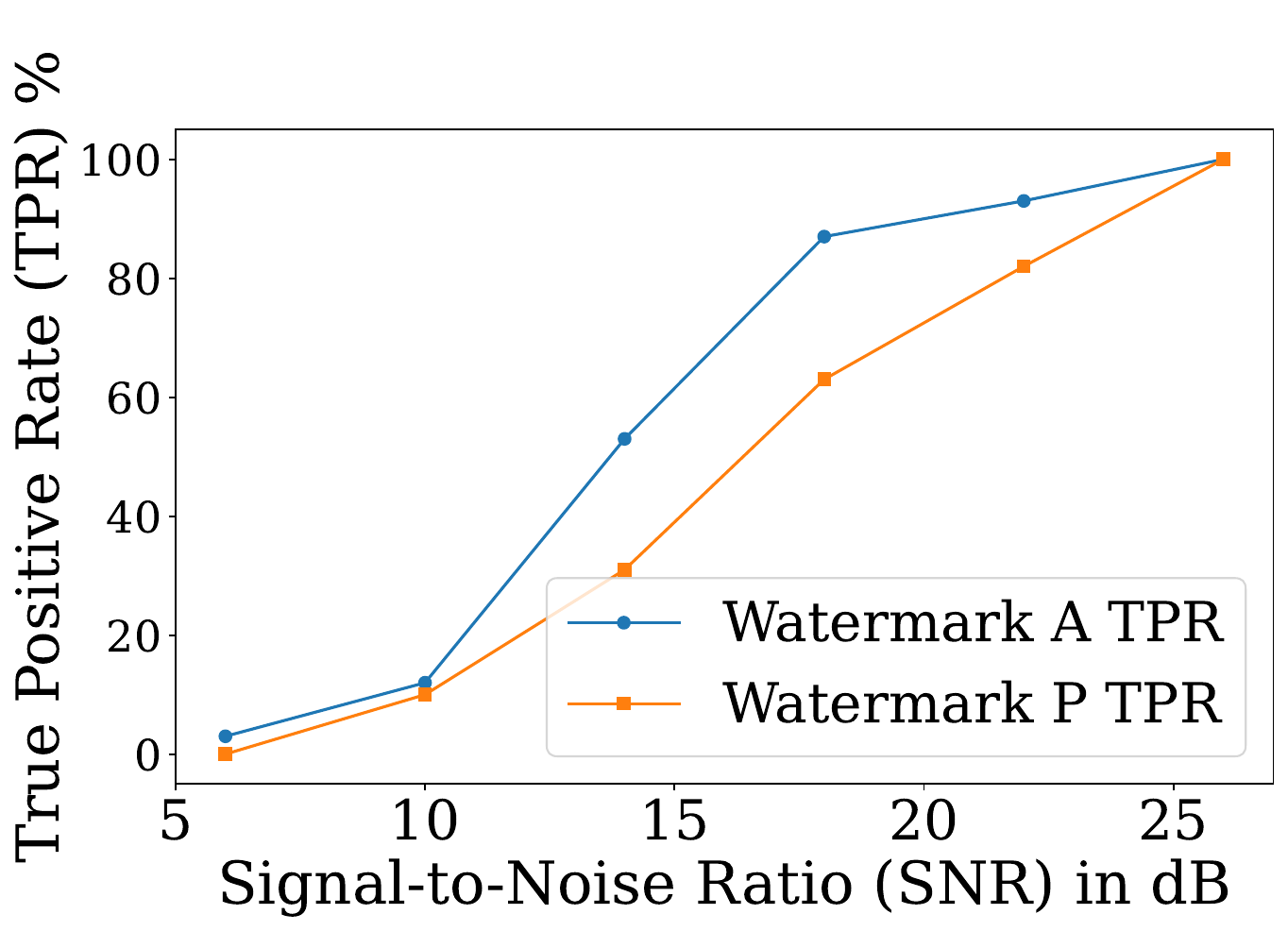}
    \centerline{(b) Room impulse response}
  \end{minipage}
  \caption{Attack-strength sweeps for A and P show nonidentical degradation under Gaussian noise and reverberation, motivating complementary evidence retention.}
  \label{fig:complementarity}
\end{figure}

Figure~\ref{fig:complementarity} shows that A and P have different relative strengths across Gaussian noise and reverberation. Multiplexing should therefore retain heterogeneous evidence rather than assume one constituent is uniformly dominant. %

\subsubsection{Constituent-Level Detection}
Table~\ref{tab:constituents_cr} reports mean TPRs over the 14 evaluation conditions under the separate-detector protocol following Eq.~\ref{eq:any_explicit}. Relative to standalone A and P in Table~\ref{tab:robustness}, parallel composition, both sequential orders, and PA-TFM offer no constituent-level mean improvement. Sequential degradation is order-dependent. MaskNet(A,P) improves both means while keeping the constituent embedders and detectors frozen.

Any-rule gains need not improve either constituent's mean detection rate: routing can retain complementary outcomes so one mark supplies evidence when the other fails. The marginal means do not quantify failure overlap or establish simultaneous survival or payload recovery; the benefit concerns evidence retention, not mere co-embedding.

\subsection{Fidelity and Controlled Analysis}
Table~\ref{tab:quality} reports pre-attack fidelity at native operating points. Unlike direct A+P addition, PA-TFM and MaskNet(A,P) recover waveform fidelity while retaining perceptual quality and ASR performance. WER remains at or below 0.04 and ABX scores are near chance, suggesting limited perceptual separability rather than proving imperceptibility.

The five quality measures remain complementary: low SNR can coexist with high PESQ/STOI, and the highest SNR does not imply the best perceptual or ASR performance. Nor is high SNR sufficient if achieved by suppressing one constituent and forfeiting complementarity.

\begin{table}[t]
\centering
\caption{Embedding fidelity before attack. WER is a fraction; ABX is closest to chance at 50\%.}
\label{tab:quality}
\scriptsize
\setlength{\tabcolsep}{3.20pt}
\renewcommand{\arraystretch}{1.08}
\begin{tabular*}{\columnwidth}{@{\extracolsep{\fill}}lccccc@{}}
\toprule
Method       & SNR$\uparrow$ & PESQ$\uparrow$ & STOI$\uparrow$ & WER$\downarrow$ & ABX (\%) \\
\midrule
A            & 25.38           & 4.45            & 1.00            & 0.01            & 54\%     \\
P            & 15.76           & 4.38            & 0.99            & 0.03            & 55\%     \\
S            & 34.75           & 4.07            & 0.99            & 0.04            & 47\%     \\
T            & 28.66           & 3.81            & 0.99            & 0.03            & 59\%     \\
W            & 36.63      & 4.08      & 0.99      & 0.01      & 50\% \\
Parallel A+P & 15.20           & 4.37            & 0.99            & 0.01            & 56\%     \\
PA-TFM       & 24.20           & 4.41            & 1.00            & 0.02            & 52\%     \\
MaskNet(A,P) & 25.36      & 4.43       & 1.00       & 0.02       & 52\%     \\
$A_\text{T}$        & 19.67      & 4.45       & 1.00       & 0.03       & 49\%     \\
MaskNet($A_\text{T}$,P) & 25.41  & 4.39       & 1.00       & 0.02       & 54\%     \\
\bottomrule
\end{tabular*}
\end{table}
Table~\ref{tab:ablation} compares the routing and proxy objectives. Removing quiet loss lowers both mean detection and SNR, while detector-aware proxy training improves ST detection over waveform-only reconstruction. These comparisons support the respective objectives but do not establish causal interference reduction or improvement of each detector.

\begin{table}[t]
\centering
\caption{Method analysis. Overall averages 14 evaluation conditions; ST denotes SpeechTokenizer. Bold: best within each block.}
\label{tab:ablation}
\scriptsize
\setlength{\tabcolsep}{5pt}
\renewcommand{\arraystretch}{1.08}
\begin{tabular*}{\columnwidth}{@{\extracolsep{\fill}}lccc@{}}
\toprule
Variant                & Overall$\uparrow$ & ST$\uparrow$ & SNR$\uparrow$ \\
\midrule
Parallel A+P           & 0.75               & 0.01          & 15.20          \\
PA-TFM                 & 0.75               & 0.01          & 24.20          \\
MaskNet w/o quiet      & 0.77               & \textbf{0.20}          & 15.34          \\
MaskNet(A,P)           & \textbf{0.88}               & \textbf{0.20}          & \textbf{25.36}          \\
\midrule
Original A             & 0.71               & 0.01          & \textbf{25.38}          \\
Waveform-only proxy    & 0.71         & 0.08    & 19.96     \\
Detector-aware $A_\text{T}$   & \textbf{0.74}               & \textbf{0.63}          & 19.67          \\
\bottomrule
\end{tabular*}
\end{table}

\subsection{SpeechTokenizer-Aware Case Study}
The released systems share a pronounced SpeechTokenizer weakness at the strict operating point, even when some of them remain robust to conventional codecs. The adapted $A_\text{T}$ markedly improves detection after the true SpeechTokenizer while retaining clean-condition detection and high perceptual quality. The gain is measured through the discrete tokenizer, not through ProxyNet.
Adaptation does not improve every condition, so $A_\text{T}$ is treated as a constituent rather than a replacement. Its AW performance decreases relative to A and it is weaker under OP and RV, even as ST improves. This creates a complementary profile that another independently detected watermark can cover.

MaskNet($A_\text{T}$,P) preserves most of the tokenizer-specific gain while restoring complementary coverage under attacks where $A_\text{T}$ alone is weak. Its fidelity remains comparable to the other high-quality settings. A channel-adapted constituent can therefore enter the learned routing framework without changing its detector interface. The evidence remains limited to one tokenizer, one adapted constituent, and two pair-specific MaskNet training runs; it does not establish universal tokenizer robustness, arbitrary-pair generalization, or security against a jointly adaptive attack.

\section{Conclusion}
We studied neural audio watermark multiplexing as allocation among independently trained systems with native detectors. The scoped benchmark reveals attack-dependent trade-offs without a stable advantage from direct composition or fixed routing. On AudioSeal–PerTh, MaskNet improves detection and fidelity, while a tokenizer-adapted constituent mitigates a shared reconstruction failure within the same framework. These findings support complementary evidence retention under the Any rule, not simultaneous mark survival, payload recovery, or improvement of every constituent. Broader pair and tokenizer generalization, payload-level coexistence, and jointly adaptive attacks remain open.

\newpage
\bibliographystyle{IEEEtran}
\bibliography{refs}

\begin{thebibliography}{10}
\providecommand{\url}[1]{#1}
\csname url@samestyle\endcsname
\providecommand{\newblock}{\relax}
\providecommand{\bibinfo}[2]{#2}
\providecommand{\BIBentrySTDinterwordspacing}{\spaceskip=0pt\relax}
\providecommand{\BIBentryALTinterwordstretchfactor}{4}
\providecommand{\BIBentryALTinterwordspacing}{\spaceskip=\fontdimen2\font plus
\BIBentryALTinterwordstretchfactor\fontdimen3\font minus
  \fontdimen4\font\relax}
\providecommand{\BIBforeignlanguage}[2]{{%
\expandafter\ifx\csname l@#1\endcsname\relax
\typeout{** WARNING: IEEEtran.bst: No hyphenation pattern has been}%
\typeout{** loaded for the language `#1'. Using the pattern for}%
\typeout{** the default language instead.}%
\else
\language=\csname l@#1\endcsname
\fi
#2}}
\providecommand{\BIBdecl}{\relax}
\BIBdecl

\bibitem{AudioSeal2024}
R.~S. Roman, P.~Fernandez, H.~Elsahar, A.~D{\'e}fossez, T.~Furon, and T.~Tran,
  ``Proactive detection of voice cloning with localized watermarking,'' in
  \emph{Proc. ICML}, 2024, pp. 43\,180--43\,196.

\bibitem{WavMark2023}
G.~Chen, Y.~Wu, S.~Liu, T.~Liu, X.~Du, and F.~Wei, ``{WavMark}: Watermarking
  for audio generation,'' \emph{arXiv preprint arXiv:2308.12770}, 2023.

\bibitem{Singh2024SilentCipher}
M.~K. Singh, N.~Takahashi, W.-H. Liao, and Y.~Mitsufuji, ``Silentcipher: Deep
  audio watermarking,'' in \emph{Proc. INTERSPEECH}, 2024, pp. 2235--2239.

\bibitem{Liu2024TimbreWatermarking}
C.~Liu, J.~Zhang, T.~Zhang, X.~Yang, W.~Zhang, and N.~Yu, ``Detecting voice
  cloning attacks via timbre watermarking,'' in \emph{Proc. NDSS}, 2024.

\bibitem{Boney1996}
L.~Boney, A.~H. Tewfik, and K.~N. Hamdy, ``Digital watermarks for audio
  signals,'' in \emph{Proc. ICMCS}, 1996, pp. 473--480.

\bibitem{Chen2001QIM}
B.~Chen and G.~W. Wornell, ``Quantization index modulation: A class of provably
  good methods for digital watermarking and information embedding,'' \emph{IEEE
  Trans. Inf. Theory}, vol.~47, no.~4, pp. 1423--1443, 2001.

\bibitem{Perth2023}
{Resemble AI}, ``Perth, a sophisticated deep neural network watermarker,''
  Software release, 2023.

\bibitem{OReilly2024MaskMark}
P.~O’Reilly, Z.~Jin, J.~Su, and B.~Pardo, ``Maskmark: Robust neural
  watermarking for real and synthetic speech,'' in \emph{Proc. ICASSP}, 2024,
  pp. 4650--4654.

\bibitem{Li2024IDEAW}
P.~Li, X.~Zhang, J.~Xiao, and J.~Wang, ``Ideaw: Robust neural audio
  watermarking with invertible dual-embedding,'' in \emph{Proc. EMNLP}, 2024,
  pp. 4500--4511.

\bibitem{NEURIPS2024_AudioMarkBench}
H.~Liu, M.~Guo, Z.~Jiang, L.~Wang, and N.~Z. Gong, ``Audiomarkbench:
  Benchmarking robustness of audio watermarking,'' in \emph{Proc. NeurIPS},
  2024.

\bibitem{ozer25_interspeech}
Y.~Özer, W.~Choi, J.~Serrà, M.~K. Singh, W.-H. Liao, and Y.~Mitsufuji, ``{A
  Comprehensive Real-World Assessment of Audio Watermarking Algorithms: Will
  They Survive Neural Codecs?}'' in \emph{{Proc. INTERSPEECH}}, 2025, pp.
  5113--5117.

\bibitem{wen2025sokrobustaudiowatermarking}
Y.~Wen, A.~Innuganti, A.~B. Ramos, H.~Guo, and Q.~Yan, ``Sok: How robust is
  audio watermarking in generative ai models?'' \emph{arXiv preprint
  arXiv:2503.19176}, 2025.

\bibitem{wu25i_interspeech}
C.-H. Wu, W.~Ge, X.~Wang, J.~Yamagishi, Y.~Tsao, and H.-M. Wang, ``{A
  Comparative Study on Proactive and Passive Detection of Deepfake Speech},''
  in \emph{Proc. INTERSPEECH}, 2025, pp. 5328--5332.

\bibitem{o'reilly2025deep}
P.~O'Reilly, Z.~Jin, J.~Su, and B.~Pardo, ``Deep audio watermarks are shallow:
  Limitations of post-hoc watermarking techniques for speech,'' in \emph{Proc.
  1st Workshop on Generative AI Watermarking at ICLR}, 2025.

\bibitem{yao2025mineoverwritingattacksneural}
L.~Yao, C.~Huang, S.~Wang, J.~Xue, H.~Guo, J.~Liu, P.~Lin, T.~Ohtsuki, and
  M.~Pan, ``Yours or mine? overwriting attacks against neural audio
  watermarking,'' \emph{arXiv preprint arXiv:2509.05835}, 2025.

\bibitem{Pavlovic2022DNNWatermark}
K.~Pavlovi{\'c}, S.~Kova{\v{c}}evi{\'c}, I.~Djurovi{\'c}, and A.~Wojciechowski,
  ``Robust speech watermarking by a jointly trained embedder and detector using
  a dnn,'' \emph{Digit. Signal Process.}, vol. 122, p. 103381, 2022.

\bibitem{Liu2023DeAR}
C.~Liu, J.~Zhang, H.~Fang, Z.~Ma, W.~Zhang, and N.~Yu, ``Dear: A
  deep-learning-based audio re-recording resilient watermarking,'' in
  \emph{Proc. AAAI}, 2023, pp. 13\,201--13\,209.

\bibitem{Li2024DRAWDR}
B.~Li, J.~Chen, Y.~Xu, W.~Li, and Z.~Liu, ``Draw: Dual-decoder-based robust
  audio watermarking against desynchronization and replay attacks,'' \emph{IEEE
  Trans. Inf. Forensics Secur.}, vol.~19, pp. 6529--6544, 2024.

\bibitem{Xu2025WAKE}
Y.~Xu, J.~Yu, H.~Chen, Z.~Wu, X.~Wu, D.~Yu, R.~Gu, and Y.~Luo, ``Wake:
  Watermarking audio with key enrichment,'' in \emph{Proc. INTERSPEECH}, 2025,
  pp. 5093--5097.

\bibitem{Liu2025XAttnMark}
Y.~Liu, L.~Lu, J.~Jin, L.~Sun, and A.~Fanelli, ``Xattnmark: Learning robust
  audio watermarking with cross-attention,'' in \emph{Proc. ICML}, 2025, pp.
  38\,987--39\,015.

\bibitem{Ji2024DiscreteWM}
S.~Ji, Z.~Jiang, J.~Zuo, M.~Fang, Y.~Chen, T.~Jin, and Z.~Zhao, ``Speech
  watermarking with discrete intermediate representations,'' in \emph{Proc.
  AAAI}, 2025, pp. 24\,239--24\,247.

\bibitem{Roman2024LatentWO}
R.~S. Roman, P.~Fernandez, A.~Deleforge, Y.~Adi, and R.~Serizel, ``Latent
  watermarking of audio generative models,'' in \emph{Proc. ICASSP}, 2025, pp.
  1--5.

\bibitem{Liu2024GROOTGR}
W.~Liu, Y.~Li, D.~Lin, H.~Tian, and H.~Li, ``Groot: Generating robust watermark
  for diffusion-model-based audio synthesis,'' in \emph{Proc. ACM MM}, 2024,
  pp. 3294--3302.

\bibitem{Ren2025P2MarkPP}
Y.~Ren, J.~Yi, T.~Wang, J.~Tao, Z.~Lian, Z.~Wen, C.~Li, R.~Fu, Y.~Bai, and
  X.~Zhang, ``{P2Mark}: Plug-and-play parameter-level watermarking for neural
  speech generation,'' \emph{arXiv preprint arXiv:2504.05197}, 2025.

\bibitem{Zhou2024TraceableSpeechTP}
J.~Zhou, J.~Yi, T.~Wang, J.~Tao, Y.~Bai, C.~Y. Zhang, Y.~Ren, and Z.~Wen,
  ``Traceablespeech: Towards proactively traceable text-to-speech with
  watermarking,'' in \emph{Proc. INTERSPEECH}, 2024, pp. 2250--2254.

\bibitem{Cheng2024HiFiGANwWS}
X.~Cheng, Y.~Wang, C.~Liu, D.~Hu, and Z.~Su, ``Hifi-ganw: Watermarked speech
  synthesis via fine-tuning of hifi-gan,'' \emph{IEEE Signal Process. Lett.},
  vol.~31, pp. 2440--2444, 2024.

\bibitem{Li2025TriniMarkAR}
Y.~Li, W.~Liu, K.~Lin, D.~Lin, and K.~Kallas, ``{TriniMark}: A robust
  generative speech watermarking method for trinity-level traceability,''
  \emph{arXiv preprint arXiv:2504.20532}, 2025.

\bibitem{Yang2025DualMarkIM}
X.~Yang, J.~Guan, F.~Xiao, C.~Fan, H.~Liu, Q.~Zhu, D.~Xu, and Y.~Lin,
  ``Dualmark: Identifying model and training data origins in generated audio,''
  \emph{arXiv preprint arXiv:2508.15521}, 2025.

\bibitem{Cho2022AttributableWO}
Y.-S. Cho, C.~S. Kim, Y.~Yang, and Y.~Ren, ``Attributable watermarking of
  speech generative models,'' in \emph{Proc. ICASSP}, 2022, pp. 3069--3073.

\bibitem{Juvela2023CollaborativeWF}
L.~Juvela and X.~Wang, ``Collaborative watermarking for adversarial speech
  synthesis,'' in \emph{Proc. ICASSP}, 2024, pp. 11\,231--11\,235.

\bibitem{Zhou2024V2AMark}
X.~Zhang, Y.~Xu, R.~Li, J.~Yu, W.~Li, Z.~Xu, and J.~Zhang, ``{V2A-Mark}:
  Versatile deep visual-audio watermarking for manipulation localization and
  copyright protection,'' in \emph{Proc. ACM MM}, 2024.

\bibitem{Cui2025EfficientSW}
Y.~Cui, P.~Pan, L.~He, and S.~Zhao, ``Efficient speech watermarking for speech
  synthesis via progressive knowledge distillation,'' \emph{arXiv preprint
  arXiv:2509.19812}, 2025.

\bibitem{Juvela2024AudioCA}
L.~Juvela and X.~Wang, ``Audio codec augmentation for robust collaborative
  watermarking of speech synthesis,'' in \emph{Proc. ICASSP}, 2025, pp. 1--5.

\bibitem{Zhou2024WMCodecEN}
J.~Zhou, J.~Yi, Y.~Ren, J.~Tao, T.~Wang, and C.~Y. Zhang, ``Wmcodec: End-to-end
  neural speech codec with deep watermarking for authenticity verification,''
  in \emph{Proc. ICASSP}, 2025, pp. 1--5.

\bibitem{Defossez2022HighFN}
A.~D{\'e}fossez, J.~Copet, G.~Synnaeve, and Y.~Adi, ``High fidelity neural
  audio compression,'' \emph{arXiv preprint arXiv:2210.13438}, 2022.

\bibitem{NEURIPS2023_DAC}
R.~Kumar, P.~Seetharaman, A.~Luebs, I.~Kumar, and K.~Kumar, ``High-fidelity
  audio compression with improved rvqgan,'' in \emph{Proc. NeurIPS}, 2023, pp.
  27\,980--27\,993.

\bibitem{2024speechtokenizer}
X.~Zhang, D.~Zhang, S.~Li, Y.~Zhou, and X.~Qiu, ``Speechtokenizer: Unified
  speech tokenizer for speech language models,'' in \emph{Proc. ICLR}, 2024.

\bibitem{Panayotov2015}
V.~Panayotov, G.~Chen, D.~Povey, and S.~Khudanpur, ``Librispeech: an asr corpus
  based on public domain audio books,'' in \emph{Proc. ICASSP}, 2015, pp.
  5206--5210.

\bibitem{MP3ISO1993}
I.~O. for Standardization~(ISO), ``Iso/iec 11172-3: Information technology —
  coding of moving pictures and associated audio for digital storage media at
  up to about 1.5 mbit/s — part 3: Audio,'' ISO, Tech. Rep., 1993.

\bibitem{OpusRFC}
J.-M. Valin, K.~Vos, and T.~Terriberry, ``Definition of the {Opus} audio
  codec,'' RFC 6716, Internet Engineering Task Force (IETF),
  \url{https://www.rfc-editor.org/rfc/rfc6716}, 2012.

\bibitem{Rix2001}
A.~W. Rix, J.~G. Beerends, M.~P. Hollier, and A.~P. Hekstra, ``Perceptual
  evaluation of speech quality (pesq)-a new method for speech quality
  assessment of telephone networks and codecs,'' in \emph{Proc. ICASSP}, 2001,
  pp. 749--752.

\bibitem{Taal2011}
C.~H. Taal, R.~C. Hendriks, R.~Heusdens, and J.~Jensen, ``An algorithm for
  intelligibility prediction of time--frequency weighted noisy speech,''
  \emph{IEEE Trans. Audio Speech Lang. Process.}, vol.~19, no.~7, pp.
  2125--2136, 2011.

\end{thebibliography}

\end{document}